\documentclass[superscriptaddress,column,showpacs,a4paper,
amssymb,amsmath,nobibnotes,aps,prd,longbibliography,
showkeys,
nofootinbib,notitlepage]{revtex4-1}
\usepackage{bigints}
\usepackage{graphicx}
\usepackage{float}
\usepackage{epsf}
\usepackage{bm}
\usepackage{amsmath}
\usepackage{amsfonts}
\usepackage{amssymb}
\usepackage{epstopdf}
\usepackage{natbib}
\usepackage{color}%
\providecommand{\U}[1]{\protect\rule{.1in}{.1in}}
\newcommand{\be}{\begin{equation}}
\newcommand{\ee}{\end{equation}}

\newcommand{\mincir}{\raise
-3.truept\hbox{\rlap{\hbox{$\sim$}}\raise4.truept\hbox{$<$}\ }}
\newcommand{\magcir}{\raise
-3.truept\hbox{\rlap{\hbox{$\sim$}}\raise4.truept\hbox{$>$}\ }}

\newtheorem{remark}{Remark}[section]
\usepackage{xcolor}
\usepackage{hyperref}
\hypersetup{colorlinks=true, linkcolor=blue,citecolor=magenta}

\newcommand{\R}{\mathbb{R}}
\begin{document}

\title{{
On the Geometric Meaning of General Relativity and the Foundations of Newtonian Cosmology
}}

\author{Jaume de Haro}
\email{jaime.haro@upc.edu}
\affiliation{Departament de Matem\`atiques, Universitat Polit\`ecnica de Catalunya, Diagonal 647, 08028 Barcelona, Spain}

\author{Emilio Elizalde}
\email{elizalde@ice.csic.es}
\affiliation{Institut de Ciències de l'Espai, ICE/CSIC and IEEC, Campus UAB, C/Can Magrans, s/n, 08193 Bellaterra, Barcelona, Spain}

\begin{abstract}

The geometric foundations of General Relativity are revisited, with particular attention to its gauge invariance, as a key to understanding the true nature of spacetime. Beyond the common image of spacetime as a deformable 'fabric' filling the Universe, curvature is interpreted as the dynamic interplay between matter and interacting fields; a view already emphasized by Einstein and Weyl, but sometimes overlooked in the literature. Building on these tools, a Newtonian framework is reconstructed that captures essential aspects of cosmology, showing how classical intuition can coexist with modern geometric insights. This perspective shifts the focus from substance to relationships, offering a fresh magnifying glass through which to reinterpret gravitational dynamics and the large-scale structure of the Universe. The similarities of this approach with other recent, more ambitious ones carried out at the quantum level are quite remarkable.

\end{abstract}

\vspace{0.5cm}


\maketitle
\section{Introduction}

Einstein's General Relativity (GR) \cite{Einstein1955} is fundamentally a gauge theory \cite{Kuzmin,Iftime-Stachel,Krasnov,Haro2024}: its physical content is invariant under active diffeomorphisms, i.e., smooth mappings that push points of the manifold to other points while transporting fields along them. This invariance, on top of being mathematically elegant, rises deep conceptual questions, epitomized by the hole argument \cite{Stachel2014,
Earman-Norton, Weatherall}. In regions of spacetime devoid of matter, the Einstein field equations alone cannot uniquely determine the metric: any active diffeomorphic image of a solution represents an equally valid configuration. 

{
At first glance, this non-uniqueness might appear to threaten determinism in GR. Yet, adopting a Machian perspective \cite{Mach}—where local inertial properties are fully determined by the distribution of mass-energy—this apparent tension is resolved. Within this view, all metrics related by an active diffeomorphism are physically equivalent: the apparent freedom of choice reflects a redundancy in the description, rather than a genuine indeterminacy in the dynamics.

This equivalence also highlights that spacetime should not be conceived as a pseudo-Riemannian manifold endowed with a unique, absolute metric structure. Such an interpretation would reify the metric as if it were an independent object. Instead, once a particular representative of the diffeomorphism-equivalence class has been chosen, the metric merely serves as a mathematical device for computing physical quantities. In this sense, spacetime is not an autonomous “deformable fabric,” but a network of relationships among matter, fields, and geometry. Likewise, coordinates acquire meaning only through the physical entity they parametrize (given a choice of metric), rather than as labels of an absolute background, thereby underscoring the relational character of the theory.

}

Building on this conceptual foundation, we will then explore how intrinsic geometric methods can be extended to a Newtonian framework, providing a reinterpretation of cosmology that remains faithful to Machian principles. Starting from perturbed Friedmann equations \cite{HEP}, which in the linear approximation coincide with the relativistic equations in the Newtonian gauge (see, for instance, \cite{Mukhanov:2005sc,Weinberg2008}), we employ tools from differential geometry—such as the Hodge operator, the Lie derivative, and differential forms—to construct a fully geometric formulation of Newtonian cosmology. This approach will allow us to describe the dynamics of matter and fields in a relational, coordinate-independent manner. Even within the fixed FLRW background, the evolution of cosmological perturbations can be cast entirely in terms of geometric and physical interactions, rather than relying on an absolute spacetime structure. In doing so, the formalism clarifies how classical Newtonian cosmology retains an underlying relational structure compatible with relativistic invariance, thus bridging the intuitive understanding of Newtonian dynamics with the more abstract, gauge-informed language of GR. Ultimately, this framework highlights the continuity between Newtonian and relativistic cosmology, showing that gauge principles and relational thinking are not merely formal abstractions but provide useful insight into the interpretation of spacetime and cosmic evolution.

\

Throughout this work we shall adopt natural units, setting $\hbar = c = k_\mathrm{B} = 1$, and use $G$ for Newton’s gravitational constant.

\section{Conceptual Meaning of the General Theory of Relativity}

Einstein’s General Theory of Relativity can be viewed as a natural extension of the Special Theory of Relativity \cite{Einstein1905}, whose original formulation did not take gravity into account. Special relativity already unified space and time into a single four-dimensional entity, spacetime (an idea due to Hermann Minkowski), but it was still a flat arena where matter moved without altering the stage on which physical phenomena occurred. General relativity transcends this limitation by introducing a geometrical description of gravity: rather than being a force in the Newtonian sense, gravity manifests itself as curvature of the spacetime.

From a mathematical standpoint, spacetime is modeled as a pseudo-Riemannian (Lorentzian) manifold \cite{O'Neill}. More precisely, it is described by a pair $({\mathcal M}, \mathfrak{g})$, where:

\begin{itemize}
\item $\mathcal{M}$ is a smooth, four-dimensional differentiable manifold, providing the underlying topological and differentiable structure—the “points” and the possible coordinate systems of spacetime.
\item $\mathfrak{g}$ is a non-degenerate, symmetric $(0,2)$ tensor field, the metric, with signature $(3,1)$, meaning that it has three positive directions corresponding to space and one negative direction corresponding to time.
\end{itemize}

This pair $({\mathcal M}, \mathfrak{g})$ contains all the essential information about the geometry of the universe: it determines how distances and time intervals are measured, how light cones define causal relations, and how free-falling particles trace out geodesics—the natural “straightest” possible paths in a curved spacetime. In this way, general relativity shifts the perspective from viewing spacetime as a static stage to recognizing it as a dynamical entity, one that interacts with energy and momentum and evolves according to Einstein’s field equations \cite{Einstein1955}:
\begin{eqnarray}
   \frak{Ric}_{\frak{g}} - \frac{1}{2} \frak{g} R_{\frak{g}} = 8\pi G \frak{T}_{\frak{g}}.
\end{eqnarray}
Here,  
$\frak{T}_{\frak{g}}$ is the stress-energy tensor representing the matter content,
$\frak{Ric}_{\frak{g}}$ is the Ricci tensor, which depends on the metric, and
$R_{\frak{g}}$ is the Ricci scalar, i.e., the trace of the Ricci tensor.

This equation has several important properties:

\begin{enumerate}
\item \textbf{General covariance:} It is built from intrinsic geometric quantities of the manifold, namely tensors, so its form remains invariant under arbitrary coordinate transformations (passive diffeomorphisms), fulfilling the Principle of Covariance \cite{NORTONa}.
\item \textbf{Conservation of energy and momentum:} The Bianchi identity \cite{Bianchi},
\begin{eqnarray}
    \mathrm{div}\left(
     \frak{Ric}_{\frak{g}} - \frac{1}{2} \frak{g} R_{\frak{g}} \right) = 0,
\end{eqnarray}
together with the field equations, implies
\begin{eqnarray}
    \mathrm{div}(\frak{T}_{\frak{g}}) = 0,
\end{eqnarray}
where the divergence of a four vector field ${\bf u}$ is defined as the trace  of the endomorphism that maps any four-vector ${\bf w}$ to the
covariant derivative $\nabla_{\bf w}{\bf u}$:
\begin{eqnarray}
    \mathrm{div}(\mathbf{u}) \equiv \mathrm{Tr}({\bf w} \mapsto \nabla_{\bf w}{\bf u}).
\end{eqnarray}
\item \textbf{Geodesic motion:} The trajectories of test particles follow the geodesics of the pseudo-Riemannian manifold, generalizing the concept of straight-line motion in flat spacetime.
\end{enumerate}

Although at first glance the theory appears elegantly simple and beautifully expressed in geometric terms, it conceals profound conceptual challenges. Interpreting the geometric formalism requires careful consideration of the interplay between mathematics and physics. Notably, spacetime is not a material substance that curves in a conventional sense; rather, the curvature encodes the gravitational interaction, determining how matter and light propagate. However, the way the theory is often presented in textbooks and lectures may foster the impression that spacetime is a pre-existing Lorentzian manifold, an absolute stage on which matter evolves—an interpretation that risks reintroducing, in disguised form, the Newtonian notion of absolute space and time. This is not what General Relativity asserts: the bare manifold carries no physical content until it is endowed with a metric, and even then the Einstein equations do not by themselves determine a unique solution. As we will show in the next section, because of the gauge freedom under active diffeomorphisms, many mathematically distinct metrics can represent the same physical situation. What the equations specify is therefore not a single, absolute spacetime, but an equivalence class of metrics, all describing the same relational structure. This perspective avoids treating spacetime as a deformable substance and instead regards it as a relational mathematical framework that allows us to calculate trajectories, such as geodesics.

Einstein himself never ceased to reflect critically on this point throughout his life, insisting that the metric and the “continuum” it describes are not autonomous entities but elements of a relational structure shaped by matter and energy. Yet, as the theory entered the mainstream, successive generations of physicists tended to regard the formalism as closed and final, focusing primarily on technical applications and new predictions. In this process, some of the original conceptual depth was set aside, and the relational insights that Einstein wrestled with were too often replaced by a more literal, almost ontological reading of the geometric formalism. Recognizing this subtle but crucial point is essential to avoid conflating the mathematical model with an ontological claim about spacetime as a substance. In what follows, we will explore these subtleties and clarify the non-trivial physical meaning behind Einstein’s elegant equations.

\subsection{The Hole Argument}

We consider a chart, i.e., a map from the pseudo-Riemannian manifold ${\mathcal M}$ (representing spacetime) to $\mathbb{R}^4$, which defines a coordinate system: 
\begin{eqnarray}
\psi: {\mathcal M}\longrightarrow \mathbb{R}^4, \qquad p\mapsto {\bf x}.
\end{eqnarray}

Outside the matter distribution — that is, in vacuum — the field equations in this coordinate system reduce to:
\begin{eqnarray}
    R_{\mu\nu}\!\left(g_{\alpha\beta}({\bf x}), {\bf x}\right)=0,
\end{eqnarray}
where $R_{\mu\nu}$ are the components of the Ricci tensor and $g_{\alpha\beta}$ those of the metric tensor.

Next, we perform a change of coordinates (in modern geometric language, a \emph{passive diffeomorphism}). That is, we choose another chart that maps the manifold to $\mathbb{R}^4$, thus defining a new coordinate system. Such a transformation preserves the matter configuration on the manifold, since a coordinate change — unlike an \emph{active diffeomorphism}, where the points of spacetime (events) are reassigned — leaves the manifold points fixed. In the new coordinates, we have: 
\begin{eqnarray}
{g}_{\mu\nu}({\bf x}) \neq \tilde{{g}}_{\mu\nu}(\tilde{\bf x}),
\end{eqnarray}
and, due to the invariance of the Ricci tensor under active diffeomorphisms, in the vacuum, we obtain:
\begin{eqnarray}
R_{\sigma\delta}\!\left(\tilde{g}_{\alpha\beta}(\tilde{\bf x}), \tilde{\bf x}\right)
= \frac{\partial x^{\mu}}{\partial\tilde{x}^{\sigma}}
\frac{\partial x^{\nu}}{\partial\tilde{x}^{\delta}}
R_{\mu\nu}\!\left(g_{\alpha\beta}({\bf x}), {\bf x}\right)
\;\;\Longrightarrow\;\;
R_{\sigma\delta}\!\left(\tilde{g}_{\alpha\beta}(\tilde{\bf x}), \tilde{\bf x}\right)=0,
\end{eqnarray}
where the functions $R_{\sigma\delta}$ are the same, i.e., no tilde appears on them.

Therefore, following Einstein's reasoning (see for example \cite{Davis1}), by simply relabeling the coordinates, one sees that:
\begin{eqnarray}
R_{\mu\nu}\!\left(\tilde{g}_{\alpha\beta}({\bf x}), {\bf x}\right)=0,
\end{eqnarray}
which shows that in the \emph{same} coordinate system, the uniqueness of the solutions is lost.
In this sense, Einstein’s procedure can be interpreted as the application of an \emph{active diffeomorphism} on the manifold, i.e., a smooth reassignment of spacetime points (events), which generates physically equivalent but mathematically distinct metric configurations.

\medskip

To put Einstein's argument in a deeper geometric context, let $\phi$ be an active diffeomorphism, i.e., a smooth map from ${\mathcal M}$ to itself. This diffeomorphism induces the pull-back of any $2$-form $\mathfrak{h}$, denoted $\phi^*\mathfrak{h}$, defined as follows: given two vector fields $X,Y \in \mathfrak{X}(\mathcal{M})$ (the set of vector fields on ${\mathcal M}$),
\begin{eqnarray}
\phi^*\mathfrak{h}(X,Y)=\mathfrak{h}(\phi_*X,\phi_*Y),
\end{eqnarray}
where $\phi_*Z$ denotes the push-forward of a vector field $Z$. The latter is defined in the usual way as
\begin{eqnarray}
\phi_*Z(f)=Z(\phi^*f)\equiv Z(f\circ \phi), \qquad \forall f\in C^\infty({\mathcal M}).
\end{eqnarray}

The crucial property is that active diffeomorphisms preserve the Ricci tensor and the scalar curvature:
\begin{eqnarray}
\phi^*\frak{Ric}_{\mathfrak{g}} = \mathfrak{Ric}_{\phi^*\mathfrak{g}}, 
\qquad 
\phi^*R_{\mathfrak{g}}=R_{\phi^*\mathfrak{g}}.
\end{eqnarray}

Therefore, applying an active diffeomorphism to the Einstein field equations yields
\begin{eqnarray}
    \mathfrak{Ric}_{\phi^*\mathfrak{g}} - \frac{1}{2} \phi^*\mathfrak{g}\, R_{\phi^*\mathfrak{g}} 
    = 8\pi G\, \phi^*\mathfrak{T}_{\mathfrak{g}},
\end{eqnarray}
where, in general, $\phi^*\mathfrak{T}_{\mathfrak{g}} \neq \mathfrak{T}_{\phi^*\mathfrak{g}}$. 
The equality holds only for the so-called \emph{active hole diffeomorphisms}, which reassign events only outside the matter distribution, leaving the matter content fixed in spacetime. Under the action of such diffeomorphisms the field equations remain unaltered. Thus, given a metric solution $\mathfrak{g}$ and a hole diffeomorphism $\phi$, the metric $\phi^*\mathfrak{g}$ is another (mathematically distinct) solution, which must nevertheless be interpreted as physically equivalent to the original one.

Finally, given two charts $\psi$ and $\tilde{\psi}$ related, respectively, with the coordinates ${\bf x}$ and $\tilde{\bf x}$, we define:
\begin{eqnarray}
    {\bf F}:\R^4\longrightarrow\R^4, \quad {\bf F(x)}
    =(\tilde{\psi}\circ \psi^{-1})({\bf x}),
\end{eqnarray}
which, given a point ${\bf p}\in {\mathcal M}$, 
sends 
${\bf x}=\psi({\bf p})$ to $\tilde{\bf x}=\tilde{\psi}({\bf p})$. Then,  the active diffeomorphism
\begin{eqnarray}
\phi: \mathcal{M}\longrightarrow
\mathcal{M}; \quad \phi=\psi^{-1}\circ {\bf F}\circ \psi=\psi^{-1}\circ \tilde{\psi},
\end{eqnarray}
 is the one associated to the Einstein's procedure.

\

Now, we review
the example of the Schwarzschild solution, which is very illuminating \cite{Macdonald}.
In polar coordinates the {Schwarzschild} (in fact, this is a solution derived by J. {Droste} \cite{Droste} and D. Hilbert \cite{Hilbert}) line element is:
\begin{eqnarray}\label{aaa}
   {ds^2= \left( 1-\frac{2MG}{r}\right)dt^2-
    \left( 1-\frac{2MG}{r}\right)^{-1}dr^2-r^2 d\Omega^2,}
\end{eqnarray}
where $d\Omega^2 = d\theta^2 + \sin^2 \theta\, d\phi^2$.

\begin{remark}
Note that the line element (\ref{aaa}) can be simply obtained by imposing that the worldline of a radially free-falling test particle satisfies Newton's second law when the coordinate time is replaced by the proper time. Effectively, we look for a line element of the form
\begin{eqnarray}\label{Hilbert_metric}
ds^2 = A(r)\, dt^2 - B(r)\, dr^2 - r^2 d\Omega^2.
\end{eqnarray}

Then, the geodesic equation for a radially free-falling test particle is
\begin{eqnarray}
-2 B' \dot{r}^2 - 2 B \ddot{r} = A' \dot{t}^2 - B' \dot{r}^2,
\end{eqnarray}
and using that 
\begin{eqnarray}
A \dot{t}^2 - B \dot{r}^2 = 1,
\end{eqnarray}
we obtain
\begin{eqnarray}
2 B \ddot{r} = -\frac{A'}{A} - \frac{(A B)'}{A} \dot{r}^2.
\end{eqnarray}

Solving for the term proportional to $\dot{r}^2$ implies $AB = \text{constant}$, and by rescaling the $r$ coordinate, we can choose $AB = 1$. Then, the geodesic equation simplifies as
\begin{eqnarray}
\ddot{r} = -\frac{A'}{2}.
\end{eqnarray}

Therefore, in order to recover Newton's second law and the Minkowski metric at large $r$, one must choose
\begin{eqnarray}
A(r) = 1 - \frac{2MG}{r}.
\end{eqnarray}
\end{remark}

Define a coordinate change $r=f(\tilde{r})$
where $r=\tilde{r}$ inside (for example) a half of  the {Schwarzschild} radius $r=2MG$, and ${f}$ is not the identity outside it. In the new coordinates the line element changes. Replacing in the new metric the variable ${\tilde{r}}$ by ${r}$, one gets
\begin{eqnarray}
   {ds^2= \left( 1-\frac{2MG}{f(r)}\right)dt^2-
    \left( 1-\frac{2MG}{f(r)}\right)^{-1}(f')^2(r)dr^2-f^2(r) d\Omega^2.}
\end{eqnarray}
As a consequence, there are different solutions to the field equations in the same coordinate system. 
However, they just correspond to different rearrangements of the events in the manifold.

For example, for the original {Schwarzschild} solution \cite{Schwarzschild}, we have 
{$f(r)=(r^3+(2MG)^3)^{1/3}$}. 
Here we can see that, in his original paper,  the {Schwarzschild} radius is translated to {$r=0$}, where there is the unique singularity of the metric. 
Both metrics are physically equivalent, yet the meaning of the coordinates is intrinsically tied to the specific choice of metric. In other words, coordinates acquire physical significance only through the way the metric is expressed.

\

This point was clearly emphasized by Misner in \cite{Misner}:

\begin{quote}
"{\it The metric defines not only the gravitational field that is assumed,
but also the coordinate system in which it is presented. There is
no other source of information about the coordinates apart from the
expression for the metric. It is also not possible to define the coordinate system in any way that does not require a unique expression
for the metric. In most cases where coordinates are chosen for
computational convenience, the expression for the metric is the most
efficient way to communicate clearly the choice of coordinates being made.}"
\end{quote}

\

The final lesson here is that the invariance of Einstein's equations under active diffeomorphisms leads to mathematically distinct solutions (i.e., different metrics) that, if one maintains the Machian spirit—where the matter content of the universe determines the inertial structure—must be considered physically equivalent. In this sense, General Relativity can be viewed as a {\it gauge theory}.

Moreover, this gauge symmetry implies that spacetime points have no physical meaning independently of the fields defined on the manifold; they do not possess an ontological existence. Consequently, one cannot treat spacetime merely as a pseudo-Riemannian manifold represented by a pair $({\mathcal M}, \frak{g})$, because any active diffeomorphism of ${\mathcal M}$ generates a mathematically distinct solution of the field equations (a different pseudo-Riemannian manifold) that is physically equivalent to the original one. Therefore, to perform specific calculations, such as determining the geodesic followed by a test particle, one must fix a gauge by selecting a particular solution (metric) of the field equations, which then allows the coordinates to acquire physical meaning relative to that chosen metric.


\subsection{
Historical Path and Conceptual Meaning of General Relativity
}

{

In his work towards the final formulation of General Relativity, Einstein followed a methodological route that differed markedly from the tradition of classical physics. In classical physics, one usually begins with a clear set of physical principles and then seeks the mathematical tools best suited to express them. By contrast, beginning around 1913 and under the decisive influence of his friend Marcel Grossmann, Einstein immersed himself in the mathematical language of Riemannian geometry and tensor calculus even before he possessed a complete and consistent physical picture of gravitation. This methodological inversion was unprecedented: whereas Newton had postulated universal gravitation from empirical regularities and only later adopted the calculus as a convenient tool, Einstein allowed mathematics itself to guide the search for the correct physical theory.


{

The years 1913–1915 are often described as the “Entwurf period,” during which Einstein and Grossmann developed a provisional theory of gravitation, the so-called Entwurf equations \cite{Grossmann}. These equations were not generally covariant, largely due to Einstein's early belief that gravitation primarily affected clock rate while leaving spatial flatness intact. This standpoint reflected his initial hesitation and conceptual struggle with the physical meaning of covariance and coordinate systems. 
As Renn and Sauer have emphasized \cite{RennSauer1999}, Grossmann’s contribution was crucial in introducing the methods of Riemannian geometry, but at this stage neither Einstein nor Grossmann fully distinguished between a Riemannian and a Lorentzian manifold, treating the invariant line element 
$ds$ mainly in analogy with Euclidean differential geometry rather than explicitly as proper time in Minkowski’s sense \cite{Minkowski}. In the Entwurf theory, Einstein wrote that $ds$ is the “{\it naturally measured distance between two space-time points}”, reflecting its formal role as a geometrical invariant rather than as a measure of time experienced by a physical clock. Janssen has similarly underlined that, in this framework, the metric was conceived first of all as a mathematical structure, only later acquiring a clear physical interpretation as the spacetime interval \cite{Janssen2005}.

}

During this phase, Einstein oscillated between two perspectives: on the one hand, his earlier physical intuitions from 1907–1912 \cite{Einstein1907,Einstein1911,Einstein1912a,Einstein1912b}, grounded in the Equivalence Principle and the analogy between gravitation and accelerated frames; on the other, the formal demands of tensor calculus, which seemed to point towards a far more general theoretical structure.

This tension culminated in late 1915, when Einstein finally abandoned the restrictions of the Entwurf theory and embraced full general covariance, arriving at the field equations in their definitive form \cite{Einstein1915}. Once these equations were written down, Einstein’s task was reversed: he now had to interpret them physically, to understand what the mathematics was telling him about the nature of space, time, and gravitation. In this sense, the development of General Relativity represents a radical departure from the methodological tradition of classical physics, marking a historical moment where the interplay between mathematical form and physical meaning became the driving force of theoretical innovation.

}

Once the final version of General Relativity was established, the first step was to choose an appropriate gauge. That is, one must select a particular solution and correctly interpret the meaning of the coordinates. For example, in the weak-field limit, where the equations can be linearized, the so-called harmonic gauge leads to a metric that generalizes Newtonian mechanics and provides a clear interpretation of the coordinates.

{

In this way, the view of spacetime emerges as a pseudo-Riemannian manifold equipped with a metric that determines distances, causal relations, and geodesic motion. Crucially, however, this ``curved spacetime'' is not a physical substance or medium that can be perceived, like Gauss's curved surfaces. Unlike the absolute space and time of classical physics, here spacetime is a \emph{relational framework}: it acquires meaning only through the network of events, trajectories, and gravitational interactions that take place within it.

Thus, ``spacetime curvature'' should not be thought of as the bending of a material sheet, but rather as a property of the relations among geodesics—that is, the way in which free-falling trajectories are organized. More precisely, matter and energy influence measurements of space and time by modifying the line element $ds^2$, the first fundamental form of the pseudo-Riemannian manifold, whose extremization defines geodesic motion. In this sense, curvature is not the deformation of a medium but the encoding of gravitational interaction within the structure of the line element itself.

At this point, it is worth recalling Weinberg’s cautionary remark \cite{Weinberg}:
\begin{quote}
{\it We may therefore express the equation of motion geometrically by saying that a particle in free fall through the curved spacetime, called a gravitational field, will move along the shortest (or longest) possible path between two points, "length" being measured by proper time. Such paths are called geodesics. For instance, we can think of the Sun as distorting spacetime just as a heavy weight distorts a rubber sheet, and we can consider a comet's path as being bent toward the Sun to keep the path as "short" as possible. However, this geometrical analogy is an a posteriori consequence of the equations of motion derived from the equivalence principle, and plays no necessary role in our considerations.}
\end{quote}

{
This sober warning is often downplayed in the pedagogical literature, where the geometrical imagery is taken more literally. In works such as those of Misner, Thorne, and Wheeler \cite{MTW}, Dirac \cite{Dirac1975}, or Hawking and Ellis \cite{HawkingEllis}, the curvature of spacetime is introduced almost as the literal substance of the theory, serving as the starting point for understanding gravitation.
Far from being a harmless metaphor, this imagery has historically distorted the very understanding of Einstein’s theory.\footnote{The now ubiquitous ``rubber sheet'' picture of spacetime curvature was not used by Einstein himself, but arose mainly in popular and pedagogical presentations of General Relativity during the mid–20th century. While visually striking, it is potentially misleading: the deformation of a two-dimensional sheet in an ambient three-dimensional space suggests a physical medium being bent, whereas the mathematical notion of spacetime curvature requires no embedding and refers only to intrinsic relations among geodesics.} 
It even misled distinguished non-specialists such as Nikola Tesla, who rejected General Relativity on precisely these grounds. In his own words \cite{Tesla}:

}

\begin{quote}
{\it
I hold that space cannot be curved, for the simple reason that it can have no properties. … To say that in the presence of large bodies, space becomes curved is equivalent to stating that something can act upon nothing. I, for one, refuse to subscribe to such a view.
}
\end{quote}

}



Tesla’s criticism illustrates the danger of confusing a pedagogical analogy with physical reality. Historically, in the Soviet Union, General Relativity faced skepticism and, in some cases, rejection precisely because its rigorous mathematical formalism—emphasizing pseudo-Riemannian manifolds, metrics, and gauge invariance—was seen as incompatible with the prevailing dialectical materialist philosophy. The theory’s abstract, relational nature conflicted with the expectation that spacetime should be a tangible, material entity. Nevertheless, some Soviet physicists, including Fock \cite{Fock}, recognized the validity of General Relativity and attempted to reconcile its concepts with dialectical materialism, striving to adapt the theory to fit the ideological framework.

From a modern, gauge-aware perspective, such “materialist” readings may be misleading: the metric is not unique, and its physical meaning emerges relationally, through interactions with matter and fields. Einstein himself explicitly addressed this point in his 1920 Leiden lecture, emphasizing that the ``ether'' of General Relativity is not a mechanical medium to be bent or stretched, but rather a set of physical conditions, encoded in the metric, which determine the motion of matter and light \cite{Einstein1920}.

\begin{quote}
{\it ``According to the general theory of relativity, space is endowed with physical qualities; in this sense, therefore, there exists an ether. According to the general theory of relativity, space without ether is unthinkable; for in such space there not only would be no propagation of light, but also no possibility of existence for standards of space and time (measuring-rods and clocks), nor therefore any spacetime intervals in the physical sense. But this ether may not be thought of as endowed with the quality characteristic of ponderable media, as consisting of parts which may be tracked through time. The idea of motion may not be applied to it.''} 
\end{quote}

This perspective aligns closely with Hermann Weyl's view on the role of geometry in physics, who emphasized that \cite{Weyl}:  

\begin{quote}
{\it The world-geometrical description is not pictorial, but rather an accurate reproduction of the state of affairs itself, so long as the concept of the continuum is understood in an abstract mathematical sense. The portrayal will be pictorial only if one replaces the number space with the space of intuition.}
\end{quote}

Weyl's statement underscores that the mathematical formalism of geometry should not be interpreted as a literal visual representation of the world, but rather as a precise, abstract tool for capturing invariants. Only when one substitutes abstract number space with intuitive spatial notions does the description become pictorial, highlighting the distinction between rigorous physical representation and intuitive visualization.

\

This viewpoint, both abstract and invariant-focused, also illuminates the conceptual transition from Special to General Relativity. In 1905, Einstein's Special Relativity \cite{Einstein1905} did not assume a pre-existing geometric unity of space and time; its focus was on the time measured by physical clocks carried along by observers and moving bodies. Later, Minkowski \cite{Minkowski} provided a geometric formulation of relativity by treating proper time,  
\[
s = \int ds,
\]  
as an invariant along worldlines. At this stage, proper time was understood purely as an invariant quantity along trajectories, without yet being incorporated as the full metric structure of a four-dimensional spacetime. It is only with the development of General Relativity that proper time becomes fully identified with the spacetime metric, unifying geometry and gravitation into a single physical framework.

\subsection{The viewpoint adopted here}

Our approach to introducing gravity in the weak limit into the infinitesimal proper time $ds$  
-as developed in our previous work \cite{Haro2025b,Haro2025a}- is rooted in clear physical principles. 
We start from the Equivalence Principle \cite{Norton}, applied to the case of a uniform gravitational potential, together with the requirement that proper time be extremized \cite{Weinberg}—by analogy with Fermat’s principle in optics \cite{Fermat}, where light follows the path that minimizes travel time.

By invoking D’Alembert’s Principle \cite{Dalembert1743}—the exact cancellation between inertial and gravitational forces—the Newtonian potential can be consistently incorporated into the infinitesimal proper time, which we regard as the essential invariant for determining trajectories. 
Next, by applying Lorentz transformations, we obtain the infinitesimal proper time $ds$ corresponding to moving matter. 
{At this stage, a crucial interpretative step is required: since we are not employing the full geometric apparatus of General Relativity, we do not interpret
$ds$ as the line element of a pre-existing curved Lorentzian manifold, but rather as the infinitesimal proper time —that is, the time measured by an ideal clock moving along with the test particle on its worldline. Although spacetime is curved, at each point one can always introduce a local inertial frame where the metric is approximately Minkowskian, so that the infinitesimal proper time coincides with the usual special-relativistic interval. Integrating these intervals along the worldline then gives the total proper time experienced by the particle, providing a clear physical meaning to $ds$ even in a curved geometry.
}

\vspace{0.2cm}

Remarkably, in the linear approximation, the resulting infinitesimal proper time, coincides with the line element of harmonic-gauge linearized GR metric:
\begin{eqnarray}
   {ds^2=(1+2\Phi)dt^2-
    8{\bf N}\cdot d{\bf x}dt
    -(1-2\Phi
 )d{\bf x}\cdot d{\bf x},}
\end{eqnarray}
where the
 {Newtonian} potential, in {Abraham's} theory \cite{Abraham}, is given by:
\begin{eqnarray}
   {\Phi({t},{\bf x})=-G\int \frac{\rho(t_{\rm r},\bar{\bf x})}{|{\bf x}-\bar{\bf x}|}d\bar{V}\Longrightarrow
     \Box\Phi=-4\pi G \rho},
\end{eqnarray}
being {$t_{\rm r}=t- |{\bf x}-\bar{\bf x}|$}  the retarded time, and 
the expression of ${\bf N}$ has to be
\begin{eqnarray}
   {{\bf N}({t},{\bf x})=-G\int \frac{\rho({t
    }_{\rm r},\bar{\bf x}){{ \bf v}}({t}_{\rm r}
    ,\bar{\bf x})}{|{\bf x}-\bar{\bf x}|}d\bar{V}}.
\end{eqnarray}

This identification provides a clear physical meaning to the harmonic gauge—a role already emphasized by Vladimir Fock \cite{Fock}. 
Thus, our construction naturally bridges a physically well motivated, principle-based derivation with the established geometric formalism.  

\vspace{0.2cm}

This interpretation, as we have already pointed out, resonates with Einstein’s own cautionary remarks: he repeatedly emphasized that geometry in relativity should not be conceived as a material fabric, but rather as a relational tool for describing how matter and fields determine geodesics. In the early stages of his work on gravitation, Einstein noted that, in the static case, gravity could be understood as a modification of the flow of time \cite{Einstein1911}. A clock situated near a massive body runs slower than one located further away, revealing that proper time itself already encodes the influence of gravity. In this sense, the impact of matter on time clearly preceded the notion of spatial curvature. Only later, with the completion of General Relativity, did the full metric formalism of curved spacetime arise as a natural generalization.



\vspace{0.2cm}

Following this original insight, our approach treats the proper time—directly determined from the Equivalence Principle and Lorentz invariance—as the physically meaningful quantity governing the motion of matter in the weak-field regime. 
This avoids the need to invoke spacetime curvature as a pre-existing substratum, and highlights instead how temporal modifications alone suffice to reproduce Newtonian gravity and its relativistic extension.  

\vspace{0.2cm}

It must be emphasized, however, that we do not actually claim this picture extends to the full, non-linear regime of General Relativity, where the curvature of the manifold becomes indispensable. 
Nevertheless, the fact that one can construct a consistent, physically grounded approximation in which proper time, rather than an abstract manifold, plays the central role, suggests that relationalism (or spacetime curvature) is not an unavoidable feature of gravitation. 
Rather, it appears only under the most generic conditions, while substantival notions of time and dynamics remain valid and physically transparent in the weak-field limit, which encompasses a large number of physical situations.

\section{Newtonian Cosmology}
\label{sec-2}

In the present section, 
we will show how, based in classical principles, one can build a cosmological theory that matches, up to linear order,  with the relativistic cosmology coming from General Relativity, where the background is based in the well-known Friedmann-Lemaître-Robertson-Walker (FLRW) metric.

\subsection{Newtonian Derivation of the Friedmann Equations}

Classical analyses such as~\cite{McCrea, Callan} have long shown that Newtonian mechanics is sufficient to capture the essential dynamics of a homogeneous, isotropically expanding, matter-dominated universe. In the domain where both Newtonian gravity and General Relativity are valid, the Newtonian description yields predictions that agree fully with the relativistic treatment of cosmic expansion.

In this section, we revisit the Newtonian derivation of the Friedmann equations, following the approach of~\cite{Ryden,deHaro24} and relying on elementary principles only, in particular, the shell theorem~\cite{Newton1687}. We consider a homogeneous ball of radius $\bar{R}$ embedded in Euclidean space, described in co-moving coordinates. Although eventually we may let $\bar{R}\to\infty$, we first assume $\bar{R}$ finite, so that the enclosed mass is
\begin{eqnarray}
\bar{M} = \frac{4\pi}{3}\,\sigma_0 \bar{R}^3,
\end{eqnarray}
with $\sigma_0$ being the (constant) mass density. The ball is assumed to expand uniformly.

Choosing the center of the ball as the origin, let ${\bf q}$ denote the coordinates of a point $P$ at some reference time, $t_0$, with normalization $a(t_0)=1$. At another time, $t$, the position of $P$ is given by $a(t){\bf q}$.

\medskip

By the shell theorem, only the mass enclosed within a radius $a(t)|{\bf q}|$ contributes to the gravitational force. A particle located at $a(t){\bf q}$ thus experiences a force given by
\begin{eqnarray}
{\bf F} = -\frac{4\pi G}{3}\,\sigma_0\,a(t){\bf q}.
\end{eqnarray}

Applying Newton’s second law to a test particle of mass $m$, yields
\begin{eqnarray}
m\,\frac{d^2}{dt^2}\!\big(a(t){\bf q}\big) = -\frac{4\pi G m}{3}\,\sigma_0\,a(t){\bf q},
\end{eqnarray}
which simplifies to the second Friedmann equation for a dust-dominated, spatially flat universe ($p_0=0$), namely
\begin{eqnarray}
\frac{\ddot{a}}{a} = -\frac{4\pi G}{3}\,\sigma_0.
\end{eqnarray}

\medskip

This result can also be obtained from a variational principle. Consider the Newtonian Lagrangian
\begin{eqnarray}
L_N = 
     \frac{\dot{a}^2}{2} + \frac{4\pi G}{3}\, a^2 \sigma_0.
\end{eqnarray}
Then, the Euler–Lagrange equation gives
\begin{eqnarray}
\ddot{a} = \frac{4\pi G}{3}\,\frac{d}{da}(a^2 \sigma_0).
\end{eqnarray}
Imposing mass conservation, namely
\begin{eqnarray}
\frac{d}{dt}\left(\frac{4\pi}{3}a^3\bar{R}^3\sigma_0 \right) =0 
\quad \Longrightarrow\quad
\frac{d}{da}(a^2\sigma_0) = -a\sigma_0,
\end{eqnarray}
we obtain, again,
\begin{eqnarray}
\ddot{a} = -\frac{4\pi G}{3}\,a\,\sigma_0.
\end{eqnarray}

\medskip

To incorporate pressure, and thus consider a general fluid, we must account for the internal energy of the particles that constitute the fluid, as well as for their interactions. This is achieved by replacing the mass density with the total energy density $\rho_0$, leading to the modified Newtonian Lagrangian~\cite{Harko}:
\begin{eqnarray}
\bar{L}_N = \frac{\dot{a}^2}{2} + \frac{4\pi G}{3}\,a^2 \rho_0.
\end{eqnarray}

Combining the Euler–Lagrange equation with the first law of thermodynamics,
\begin{eqnarray}
    d(a^3\rho_0)=-p_0\,d(a^3),
\end{eqnarray}
where $p_0$ is the uniform pressure, 
we recover the second Friedmann equation \cite{Friedmann}:
\begin{align}
\frac{\ddot{a}}{a} = -\frac{4\pi G}{3}\,(\rho_0 + 3p_0).
\end{align}

\medskip

Finally, to obtain the first Friedmann equation for a flat FLRW universe, we write the first law of thermodynamics as
\begin{eqnarray}
    \dot{\rho}_0+3H(\rho_0+p_0)=0,
\end{eqnarray}
where we have introduced the homogeneous Hubble rate $H\equiv \frac{\dot{a}}{a}$.  

Now, we write the second Friedmann equation as follows
\begin{align}
\frac{\ddot{a}}{a} = -{4\pi G}\,(\rho_0 + p_0)+\frac{8\pi G}{3}\, \rho_0,
\end{align}
and substituting $\rho_0+p_0=-\dot{\rho}_0/(3H)$, we get
\begin{align}
\frac{\ddot{a}}{a} = 
\frac{4\pi G}{3H}\,\dot{\rho}_0+\frac{8\pi G}{3}\, \rho_0.
\end{align}

Multiplying by $a\dot{a}$ leads to
\begin{eqnarray}
    \frac{1}{2}\frac{d}{dt}(\dot{a}^2)
    =\frac{4\pi G}{3}\frac{d}{dt}(a^2\rho_0),
\end{eqnarray}
which integrates to
\begin{align}
\dot{a}^2-\frac{8\pi G}{3}a^2\rho_0=C,
\end{align}
where $C$ is an integration constant related to the spatial curvature.  
For a flat universe, $C=0$, yielding the first Friedmann equation:
\begin{align}
H^2 = \frac{8\pi G}{3}\,\rho_0.
\end{align}

\subsection{Evolution of perturbations in an expanding universe in the Newtonian approach}
\label{sec-3}

In this section, we outline a framework in which the evolution of perturbations in an expanding universe can be studied, within a certain approximation, without invoking Einstein’s field equations of General Relativity. The central assumption is the conservation of the stress–energy tensor, which is equivalent to the relativistic continuity and Euler's equations. This postulate is well motivated by the Principle of Covariance \cite{Stachel1980}: since the stress–energy tensor is conserved in special relativity, particularly in a locally inertial frame, it must remain conserved in any frame, as the covariant derivative reduces to the partial derivative in local inertial coordinates, and tensor relations hold universally.  

Beyond this, we explore suitable extensions of the homogeneous Friedmann equations. These modifications are required not only to incorporate the classical Poisson equation but also to reproduce, at first order in perturbations, the Hamiltonian and momentum constraints of General Relativity \cite{Deser}.  

The constitutive equations that arise from this construction may be viewed as an enlarged version of the classical equations of motion. They are considerably simpler than the full Einstein equations, most notably lacking full general covariance, but they nonetheless capture some essential features of relativistic dynamics and satisfy standard classical tests. This approach thus provides a tractable framework for investigating the role of modified metrics, while retaining the core physical principles that connect the classical and the relativistic theories of gravity.

When both the cosmic expansion (described by the scale factor) and the perturbations of the Newtonian potential are taken into account, the line element—more precisely, the infinitesimal proper time—in the Newtonian gauge takes the form \cite{Mukhanov:2005sc}:
\begin{eqnarray}\label{cosmometric}
ds^2 = \big(1+2\Phi({\bf q}, t)\big)dt^2 - a_{\rm N}^2({\bf q}, t)\, d{\bf q}\cdot d{\bf q}.
\end{eqnarray}
Here we have introduced
\begin{eqnarray}
a_{\rm N}^2({\bf q}, t) \equiv \frac{a^2(t)}{
1+2\Phi({\bf q}, t)}\cong a^2(t)\,\big(1-2\Phi({\bf q}, t)\big),
\end{eqnarray}
so that $a_{\rm N}$ can be interpreted as the perturbed scale factor.

\subsubsection{The Perturbed Friedmann Equations}

We begin by expanding the energy density and pressure as 
$\rho=\rho_0+\delta \rho$ and $p=p_0+\delta p$, 
where $\delta$ denotes a small perturbation of the corresponding background quantity.  
In physical coordinates ${\bf x}=a(t){\bf q}$, the classical Poisson equation then reads
\begin{eqnarray}
    \Delta_{\bf x}\Phi = 4\pi G\,\delta\rho,
\end{eqnarray}
which, in terms of co-moving coordinates ${\bf q}$, can be rewritten as
\begin{eqnarray}
    -\frac{1}{3}\,\Delta_{\bf q}(-2a^2\Phi) = \frac{8\pi G}{3}\,a^4 \delta\rho.
\end{eqnarray}

To reconcile the homogeneous first Friedmann equation,
$$H^2 = \frac{8\pi G}{3}\rho_0,$$
with the Poisson equation, we adopt the following generalization of the Friedmann equation~\cite{HEP}:
\begin{eqnarray}\label{perturbedFriedmann}
    H_{\rm N}^2 - \frac{1}{3a_{\rm N}^4}\Delta_{\bf q}a_{\rm N}^2 = \frac{8\pi G}{3}\rho,
\end{eqnarray}
where the Hubble rate in {\it cosmic time} is defined as
\begin{eqnarray}
H_{\rm N} \equiv \frac{1}{a_{\rm N}}\,\partial_{\bf t}a_{\rm N}
\cong H - H\Phi - \partial_t\Phi,
\end{eqnarray}
with 
$\partial_{\bf t} = \tfrac{a_{\rm N}}{a(t)}\partial_{t}$ 
the unit time-like derivative.  

\medskip

Following a similar procedure,
in~\cite{HEP}, the two perturbed Friedmann equations are chosen so that they closely resemble the homogeneous ones:
\begin{eqnarray}
&& H_{\rm N}^2 - \frac{1}{3a_{\rm N}^4}\Delta_{\bf q}a_{\rm N}^2 = \frac{8\pi G}{3}\rho,\\
&& \frac{\partial_{\bf t}^2 a_{\rm N}}{a_{\rm N}} + \frac{1}{6a_{\rm N}^4}\Delta_{\bf q}a_{\rm N}^2 = -\frac{4\pi G}{3}(\rho+3p).
\end{eqnarray}

Since we work at linear order in perturbations, these equations can be recast in equivalent forms. For example, using the identity
\begin{align}
\Delta_{\bf q} a_{\rm N}^2 = 2\big(\nabla_{\bf q}a_{\rm N}\cdot \nabla_{\bf q}a_{\rm N} + a_{\rm N}\Delta_{\bf q}a_{\rm N}\big),
\end{align}
and neglecting the quadratic term 
$\nabla_{\bf q}a_{\rm N}\cdot \nabla_{\bf q}a_{\rm N}$,
the perturbed Friedmann equations reduce to
\begin{eqnarray}\label{Friedmanm1}
&& H_{\rm N}^2 - \frac{2}{3a_{\rm N}^3}\Delta_{\bf q}a_{\rm N} = \frac{8\pi G}{3}\rho,\\
&& \frac{\partial_{\bf t}^2 a_{\rm N}}{a_{\rm N}} + \frac{1}{3a_{\rm N}^3}\Delta_{\bf q}a_{\rm N} = -\frac{4\pi G}{3}(\rho+3p).
\end{eqnarray}

Next, introducing the {\it space-like Hubble vector}
\begin{eqnarray}
{\bf H}_{\rm N}\equiv\frac{\nabla_{\bf q}a_{\rm N}}{a_{\rm N}}
\cong -\nabla_{\bf q}\Phi,
\end{eqnarray}
and noting that
\begin{eqnarray}
\Delta_{\bf q}a_{\rm N}=\nabla_{\bf q}a_{\rm N}\cdot{\bf H}_{\rm N}
+ a_{\rm N}\,\nabla_{\bf q}\cdot {\bf H}_{\rm N},
\end{eqnarray}
we finally obtain, at first order in perturbations,
\begin{eqnarray}\label{Friedmann2}
&& H_{\rm N}^2 - \frac{2}{3a_{\rm N}^2}\nabla_{\bf q}\cdot{\bf H}_{\rm N} = \frac{8\pi G}{3}\rho,\\\label{Friedmann2a}
&& \frac{\partial_{\bf t}^2 a_{\rm N}}{a_{\rm N}} + \frac{1}{3a_{\rm N}^2}\nabla_{\bf q}\cdot{\bf H}_{\rm N} = -\frac{4\pi G}{3}(\rho+3p).
\end{eqnarray}

\medskip

It is important to observe that these equations coincide with the first-order perturbative equations of GR. Indeed, at first order, the perturbed Einstein equations read~\cite{Mukhanov:2005sc}
\begin{eqnarray}
    \Delta_{\bf q}\Phi - 3H\big(\partial_t\Phi + H\Phi\big) = 4\pi G\,\delta\rho,\\
    \partial_t^2\Phi + 4H\partial_t\Phi + (2\dot{H}+3H^2)\Phi = 4\pi G\,\delta p,
\end{eqnarray}
which, after combination, reproduce the perturbed Friedmann equations, (\ref{Friedmann2}) and (\ref{Friedmann2a}), up to linear order.


Next, we deal with the conservation of the stress tensor. Given the line element (\ref{cosmometric}), the conservation equation for a perfect fluid reads
\[
\mbox{div}(\mathfrak{T}_{\bf g})=0,
\]
 where the energy-momentum tensor takes the form:
\begin{equation}
\mathfrak{T}_{\frak{g}} = (\rho +  p) \, {\bf u}^{\flat} \otimes {\bf u}^{\flat} -  p \, \mathfrak{g},
\end{equation}
with the four-velocity being
\[
{\bf u}=(a_{\rm N}\sqrt{1+a_{\rm N}^2|{\bf v}|^2}, {\bf v}), 
\quad
{\bf v}=\frac{d{\bf q}}{ds},
\]
and ${\bf u}^{\flat}$ denotes the dual form of ${\bf u}$, i.e., 
\[
{\bf u}^{\flat}({\bf w})=\mathfrak{g}({\bf u},{\bf w}) \quad \forall {\bf w} \in \frak{X}(\mathcal{M}).
\]

At first order in ${\bf v}$, this leads to:
\begin{align}\label{Conservation_Euler}
& \partial_{\bf t}\rho+{\bf v}\cdot\nabla_{\bf q}\rho+
(\rho+p)\left[3H_{\rm N}+\nabla_{\bf q}\cdot{\bf v}+2{\bf H}_{\rm N}\cdot{\bf v}\right]=0,\\
& \partial_{\bf t}\big(a_{\rm N}^2(\rho+p){\bf v}\big)+
 (\rho+p)\big[ 3H_{\rm N}a_{\rm N}^2{\bf v}-{\bf H}_{\rm N} \big]
 +\nabla_{\bf q}p=0,\label{Euler_final}
\end{align}
where ${\bf v}\equiv \frac{d{\bf q}}{ds}$, which aligns, at first order in perturbations, with those of GR. Indeed, these equations can be written as
\begin{equation}
\left\{
\begin{aligned}
 \partial_t{\delta\rho} + (\rho_0+p_0) \left[\nabla_{\bf q} \cdot {\bf v} - 3\partial_t{\Phi}\right] + 3H(\delta\rho+\delta p) &= 0,\\
\partial_t\left( (\rho_0+p_0) a^2{\bf v} \right) + (\rho_0+p_0) \left[3Ha^2{\bf v} + \nabla_{\bf q} \Phi\right] + \nabla_{\bf q} \delta p &= 0.
\end{aligned}
\right.
\end{equation} 

The combination of the two Friedmann equations and the conservation equation leads to the {\it diffeomorphism constraint}:
\begin{align}
\Delta_{\bf q}\partial_{\bf t}a_{\rm N}=4\pi Ga_{\rm N}^3(\rho_0+p_0)\nabla_{\bf q}\cdot{\bf v}
\quad\Longrightarrow\quad
4\pi G\, a^2 (\rho_0+p_0) \, \nabla_{\bf q}\cdot{\bf v} = -\Delta_{\bf q}(\partial_t \Phi + H\Phi),
\end{align}
or, using the Helmholtz decomposition of the velocity \cite{Marsden},
\[
{\bf v} = {\bf v}_{\parallel} + {\bf v}_{\perp},
\]
where ${\bf v}_{\parallel}$ is curl-free and ${\bf v}_{\perp}$ is divergence-free, one arrives at
\begin{equation}
4\pi G\, a^2 (\rho_0 + p_0)\, {\bf v}_{\parallel} = -\nabla_{\bf q}(\partial_t \Phi + H\Phi),
\end{equation}
which corresponds to Eq.(7.48) of \cite{Mukhanov:2005sc}.

In addition, the second Friedmann equation can be written as:
\begin{align}
\label{Ricci}\partial_{\bf t}H_{\rm N}+\frac{1}{a_{\rm N}^2} \nabla_{\bf q}\cdot {\bf H}_{\rm N}=-4\pi G(\rho+p),
\end{align}
or equivalently,
\begin{align}
2\partial_{\bf t}H_{\rm N}+3H_{\rm N}^2=-8\pi G p,
\end{align}
and thus, recalling that the first Friedmann equation is a constraint, one can take as dynamical equations (\ref{Ricci}) together with (\ref{Conservation_Euler}) and (\ref{Euler_final}).

In terms of the four-gradient of a scalar function $f$, defined via the {\it sharp operator} as:
\[
\mbox{grad}(f)\equiv df^{\sharp}, \quad \text{with} \quad \mathfrak{g}(df^{\sharp}, {\bf w})=df({\bf w})
\quad \forall {\bf w} \in \frak{X}(\mathcal{M}),
\]
and the four-divergence of a four-vector ${\bf w}$, $\mbox{div} ({\bf w})$, the dynamical Friedmann equation can be written in geometric form:
\begin{equation}\label{geometric}
\mbox{div}({\bf h}_{\rm N})=4\pi G(\rho-p),
\end{equation}
where we have introduced the {\it Hubble four-vector}:
\begin{align}
{\bf h}_{\rm N}\equiv\frac{1}{a_{\rm N}}\mbox{grad}(a_{\rm N}) = H_{\rm N}\partial_{\bf t}-\frac{1}{a_{\rm N}^2}{{\bf H}}_{\rm N}.
\end{align}
A direct calculation yields
\begin{align}
\mbox{div}({\bf h}_{\rm N}) = \frac{\partial^2_{{\bf t}^2}a_{\rm N}}{a_{\rm N}} + 2H_{\rm N}^2 - \frac{1}{a_{\rm N}^2}\nabla_{\bf q}\cdot{\bf H}_{\rm N},
\end{align}
which reproduces (\ref{Friedmann2a}) plus twice (\ref{Friedmann2}), giving (\ref{geometric}).


{

Alternatively, one can introduce the {\it modified stress tensor}, which differs in sign from the tensor $ \frak{T}_{\frak{g}}-\frac{1}{2}\frak{g}T$ originally used by Einstein in his formulation of the field equations 
\cite{Einstein1915}:
\begin{eqnarray}
    \hat{\frak{T}}_{\frak{g}}\equiv \frak{T}_{\frak{g}}+\frac{1}{2}\frak{g}T,
\end{eqnarray}
where $T$ is the trace of the stress tensor, and taking into account that,
for the four velocity, one has
$\hat{\frak{T}}_{\frak{g}}({\bf u},{\bf u})=\frac{3}{2}(\rho-p)$, one arrives at the final form of the dynamical Friedmann equation:
\begin{equation}\label{geometric1}
\mbox{div}({\bf h}_{\rm N})=\frac{8\pi G}{3}\hat{\frak{T}}_{\frak{g}}({\bf u},{\bf u}).\end{equation}
}

On the other hand, given the metric (\ref{cosmometric}), 
and taking into account the conservation of the stress tensor, the full dynamical equations at first order in perturbations acquire the compact geometric form:
\begin{equation}\label{finaleq}
\mathrm{div}({\bf h}_{\rm N})=\frac{8\pi G}{3}\hat{\frak{T}}_{\frak{g}}({\bf u},{\bf u}),
\qquad
\mathrm{div}(\mathfrak{T}_{\frak{g}})=0\Longleftrightarrow
\mathrm{div}(\hat{\frak{T}}_{\frak{g}})=\frac{1}{2}dT,
\end{equation}
where the conservation of the stress-energy tensor leads to the relativistic conservation and Euler's equations \cite{Comer}:
\begin{equation}
    \frac{d \rho}{ds} = -\rho_{\rm in} \, \mathrm{div}(\mathbf{u}), \qquad
    \rho_{\rm in} \, \frac{D \mathbf{u}}{ds} = \mathrm{grad}( p) - \frac{d  p}{ds} \, \mathbf{u},
\end{equation}
where $D/ds$  denotes the covariant derivative along the worldline, parameterized by $s$, and 
$\rho_{\rm in}\equiv \rho+p$ is what
Weinberg (and also Landau and Lifshitz \cite{Landau}) call the {\it effective inertial mass density} \cite{Weinberg}.

Note that the relativistic conservation equation can also be written as
\begin{equation}
    \mathrm{div}(\rho{\bf u})+p\,\mathrm{div}({\bf u})=0 \quad \Longleftrightarrow \quad 
    \mathrm{div}((\rho+p){\bf u})=\frak{g}({\bf u}, \mathrm{grad}(p)).
\end{equation}

Therefore, the constituent equations can be put as
\begin{equation}\label{finaleq1}
\mathrm{div}({\bf h}_{\rm N})=\frac{8\pi G}{3}\hat{\frak{T}}_{\frak{g}}({\bf u},{\bf u})
\qquad
\mathrm{div}(\rho_{\rm in}{\bf u})=\frak{g}({\bf u}, \mathrm{grad}(p)),\qquad
\rho_{\rm in}\frac{D{\bf u}}{ds}=\mathrm{grad}(p)-\frac{dp}{ds}{\bf u}.
\end{equation}

These equations can be expressed in a more geometric form by noting that $\frac{dp}{ds}=\frak{g}({\bf u}, \mathrm{grad}(p))$:
\begin{equation}\label{finaleq2}
\mathrm{div}({\bf h}_{\rm N})=\frac{8\pi G}{3}\hat{\frak{T}}_{\frak{g}}({\bf u},{\bf u})
\qquad
\mathrm{div}(\rho_{\rm in}{\bf u})=\frak{g}({\bf u}, \mathrm{grad}(p)),\qquad
\rho_{\rm in}\frac{D{\bf u}}{ds}=\mathrm{grad}^{\perp}(p),
\end{equation}
where $\mathrm{grad}^{\perp}(p)\equiv
\mathrm{grad}(p)-\frak{g}({\bf u}, \mathrm{grad}(p)){\bf u}$ is the orthogonal rejection of $\mathrm{grad}(p)$ onto ${\bf u}$.
Alternatively, using the sharp operator and noting that $df({\bf w}) = \frak{g}({\bf w}, \mathrm{grad}(f))$, the dynamical equations take their simplest geometric form:
\begin{equation}\label{finaleq3}
\mathrm{div}({\bf h}_{\rm N})=\frac{8\pi G}{3}\hat{\frak{T}}_{\frak{g}}({\bf u},{\bf u})
\qquad
\mathrm{div}(\rho_{\rm in}{\bf u})=dp({\bf u}),\qquad
\rho_{\rm in}\frac{D{\bf u}}{ds}=(dp^{\sharp})^{\perp},
\end{equation}
In their most formal geometric expression, using the Hodge star operator and the Lie derivative (see, for instance, \cite{Frankel}), these equations read:
\begin{equation}\label{finaleq4}
\ast d\ast {\bf h}_{\rm N}^{\flat}=\frac{8\pi G}{3}\hat{\frak{T}}_{\frak{g}}({\bf u},{\bf u}),
\qquad
\ast d\ast\big(\rho_{\rm in}{\bf u}^{\flat}\big)=dp({\bf u}),\qquad
\rho_{\rm in} \, \bigl( (\mathcal{L}_{\bf u} {\bf u}^{\flat})^\sharp \bigr)^\perp
= (dp^\sharp)^\perp.
\end{equation}

\begin{remark}
  
In terms of the Hubble four-vector ${\bf h}_{\rm N}$, the four-velocity ${\bf u}$, and the modified stress–energy tensor $\hat{\frak{T}}$, the dynamical equations of Newtonian cosmology take the form
\begin{eqnarray}
\mathrm{div}({\bf h}_{\rm N}) = \frac{8\pi G}{3}\,\hat{\frak{T}}_{\frak{g}}({\bf u},{\bf u}), \qquad
\mathrm{div}(\hat{\frak{T}}_{\frak{g}}) = \frac{1}{6}\,d\hat{T},
\end{eqnarray}
which, in the language of differential forms, can be equivalently written as
\begin{eqnarray}
\ast d\ast {\bf h}_{\rm N}^{\flat} = \frac{8\pi G}{3}\,\hat{\frak{T}}_{\frak{g}}({\bf u},{\bf u}), \qquad
\ast d\ast \hat{\frak{T}}_{\frak{g}} = \frac{1}{6}\,d\hat{T},
\end{eqnarray}
where we have used that the trace of $\hat{\frak{T}}_{\frak{g}}$, namely $\hat{T}$, satisfies $\hat{T}=3T$.

On the other hand, defining $a_{\rm N}\equiv e^{\phi}$ and using the Laplace–Beltrami operator $\Box_{\frak{g}}$, the equations become
\begin{eqnarray}
\Box_{\frak{g}}\phi = \frac{8\pi G}{3}\,\hat{\frak{T}}_{\frak{g}}({\bf u},{\bf u}), \qquad
\mathrm{div}(\hat{\frak{T}}_{\frak{g}}) = \frac{1}{6}\,d\hat{T},
\end{eqnarray}
or, equivalently, in terms of the Hodge star operator,
\begin{eqnarray}
\ast d\ast d\phi = \frac{8\pi G}{3}\,\hat{\frak{T}}_{\frak{g}}({\bf u},{\bf u}), \qquad
\ast d\ast \hat{\frak{T}}_{\frak{g}} = \frac{1}{6}\,d\hat{T}.
\end{eqnarray}

\end{remark}

\

\begin{remark}
For a dust fluid, the dynamical equations simplify to
\begin{equation}\label{dusteq}
\mathrm{div}({\bf h}_{\rm N})=4\pi G \rho, \qquad
\mathrm{div}(\rho{\bf u})=0, \qquad
\frac{D{\bf u}}{ds}=0,
\end{equation}
showing the well-known result that a dust fluid follows the geodesics of spacetime.

In fact, for a dust fluid, instead of the conservation of the stress-energy tensor, one can use an alternative conservation law. Consider an arbitrary subset $\Omega$ of spacetime and the fluid flow $\varphi_{s}$, defined by $\frac{d\varphi_{ s}({t},{\bf q})}{ds}={\bf u}({t},{\bf q})$, with $\varphi_{s_0}$ the identity. Then, using
\begin{equation}
    \left[\frac{d}{ds}\int_{\varphi_s(\Omega)}f({t},{\bf{q}})\,dV\right]_{s=s_0}=\int_{\Omega} \mathrm{div}(f{\bf u})_{s=s_0}\,dV,
\end{equation}
where for our metric $dV=a_{\rm N}^2a(t)\,d{t}\,dq_1\,dq_2\,dq_3$, and applying "mass conservation":
\begin{equation}
    \left[\frac{d}{ds}\int_{\varphi_s(\Omega)} \rho({t},{\bf{q}})\,dV\right]_{s=s_0}=0,
\end{equation}
we obtain $\mathrm{div}(\rho {\bf u})=0$.
\end{remark}

\


To conclude this section, a few words on vector and tensor perturbations are in order.
The {\it vector perturbations}, which are independent of both the Newtonian potential and the perturbed matter content, are associated with the divergence-free component of the velocity, ${\bf v}{\perp}$. Indeed, in the absence of these perturbations, the conservation equation reduces to $\nabla_{\bf q} \cdot {\bf v} = 0$. Consequently, when matter perturbations are neglected, the relativistic Euler equation simplifies to
\begin{equation}
    \partial_t\left(
    (\rho_0 + p_0)a^2{\bf v}_{\perp}
\right)
+ 3H(\rho_0 + p_0)a^2{\bf v}_{\perp}
 = 0,
\end{equation}
implying that
\begin{equation}
    |{\bf v}_{\perp}|\propto \frac{1}{a^5(\rho_0+p_0)},
\end{equation}
so that the {\it physical} rotational velocity, $a{\bf v}_{\perp}$, decays as $1/[a^4(\rho_0+p_0)]$, which in a matter-dominated universe reduces to $1/a$, as in the Newtonian theory. Therefore, vector perturbations, denoted ${\bf S}$ and defined in GR as \cite{Mukhanov:2005sc}:
\begin{equation}
    \Delta {\bf S}=-16\pi G a^3(\rho_0+p_0){\bf v}_{\perp},
\end{equation}
do decay very quickly and can be neglected.

Next, {\it tensor perturbations}, which have no Newtonian analogue, appear in linearized GR as gravitational waves. Since they are perturbations of the FLRW spacetime, using the Laplace-Beltrami operator in this spacetime, namely $\Box_{\rm FLRW}$, they satisfy the wave equation in this background, namely
\begin{equation}
    \Box_{\rm FLRW} h_{ij}=0 \quad \Longrightarrow \quad
    \ddot{h}_{ij}+3H\dot{h}_{ij}-\frac{1}{a^2}\Delta h_{ij}=0,
\end{equation}
where $h_{ij}$ is symmetric, transverse ($\partial_i h^i_j=0$), and traceless ($h^i_i=0$).

\

\section{Conclusions}


In this work, we have explored the close connection that can be established among the nature of spacetime, gauge invariances, and the formulation of cosmology, both in relativistic and in Newtonian settings. General Relativity, understood as a gauge theory that is invariant under active diffeomorphisms, displays a well-known non-uniqueness of the metric. This is exemplified by the hole argument, as described above with care. From a Machian standpoint, however, such gauge freedom does not imply indeterminacy; rather, it actually highlights the relational character of spacetime. Namely, the fact that all metrics connected by active diffeomorphisms are physically equivalent. Thus, coordinates acquire meaning only through the physical fields they describe.

This perspective emphasizes that spacetime should not be thought of as a deformable 'fabric', but instead, as a web of relations among matter and fields. Within this framework, the central invariant that governs the dynamics of bodies is ${ds}$, which is best understood not as an abstract geometric line element, but as the infinitesimal proper time elapsed along a trajectory.


{
Most interestingly, while Einstein himself consistently maintained a deeply critical perspective on the theory he had created, later generations of physicists—figures such as Penrose, Wheeler, or Hawking—have frequently treated General Relativity as a final, completed, and immutable framework. Their focus on deriving new consequences, finding explicit solutions, or exploring applications in cosmology (such as gravitational lenses and gravitational waves) often led to the neglect of the many conceptual subtleties that Einstein considered essential. In effect, these authors sometimes became more doctrinaire than the founder himself. The heavy use of sophisticated mathematical machinery, while elegant and powerful, can obscure the underlying physical meaning: foundational questions such as the relational character of the metric, the true physical significance of coordinates, or the implications of the hole argument were frequently overlooked. Consequently, much of Einstein’s critical spirit—his insistence on interpreting and questioning the very meaning of the geometric structure—gradually receded in favor of extensive technical developments and practical applications.

}

Here, building on this conceptual foundation, we have developed a geometric formulation of Newtonian cosmology that preserves the relational structure of the theory while remaining consistent with a fixed FLRW background. By intrinsically employing geometric objects, such as the Hodge operator and Lie derivatives, our framework has been proven to achieve formal invariance under passive diffeomorphisms, thus rendering explicit the gauge-informed structure of classical cosmology. Starting from the perturbed Friedmann equations, which in the linear approximation coincide with the relativistic equations in the Newtonian gauge, we have proven that cosmological dynamics can be fully captured in terms of interactions among matter and fields, without the need to invoke an absolute background structure beyond the FLRW metric.

This approach bridges the gap between Newtonian intuition and the invariant language of General Relativity, showing that even in classical limits, a coherent geometric and relational understanding of spacetime and the cosmological evolution is truly possible. It highlights, moreover, how differential geometric tools allow a precise and gauge-aware reformulation of Newtonian cosmology, clarifying in this way the conceptual foundations of classical cosmology, and providing a stepping stone for the exploration of more general relational and Machian frameworks, beyond the standard Newtonian and relativistic descriptions.

While staying strictly at the classical level, our approach is somehow reminiscent of more ambitious attempts (regretfully not successful, as of now), carried out at a deeper quantum level, with the aim of understanding the emergence of spacetime from quantum correlations of energy-matter fields. In those approaches, it is quantum entanglement that knits the web, the 'fabric' of spacetime together. Its geometry is determined there by the entanglement pattern in the underlying quantum system. In short, the geometry of a region of spacetime emerges from the entanglement entropy of the corresponding quantum field theory. The similarities with our description of the classical level are quite remarkable. It will be worth exploring whether our results here can also be useful in this more fundamental context.

\begin{acknowledgments}

JdH is supported by the Spanish grant PID2021-123903NB-I00
funded by MCIN/AEI/10.13039/501100011033 and by ``ERDF A way of making Europe''. EE is partly supported by the program Unidad de Excelencia María de Maeztu, CEX2020-001058-M, and by AGAUR, project 2021-SGR-00171.  
\end{acknowledgments}

\end{document}